\makeatletter

\font\tenmsx=msxm10
\font\sevenmsx=msxm7
\font\fivemsx=msxm5
\font\tenmsy=msym10
\font\sevenmsy=msym7
\font\fivemsy=msym5
\newfam\msxfam
\newfam\msyfam
\textfont\msxfam=\tenmsx  \scriptfont\msxfam=\sevenmsx
  \scriptscriptfont\msxfam=\fivemsx
\textfont\msyfam=\tenmsy  \scriptfont\msyfam=\sevenmsy
  \scriptscriptfont\msyfam=\fivemsy

\def\hexnumber@#1{\ifnum#1<10 \number#1\else
 \ifnum#1=10 A\else\ifnum#1=11 B\else\ifnum#1=12 C\else
 \ifnum#1=13 D\else\ifnum#1=14 E\else\ifnum#1=15 F\fi\fi\fi\fi\fi\fi\fi}

\def\msx@{\hexnumber@\msxfam}
\def\msy@{\hexnumber@\msyfam}
\mathchardef\boxdot="2\msx@00
\mathchardef\boxplus="2\msx@01
\mathchardef\boxtimes="2\msx@02
\mathchardef\square="0\msx@03
\mathchardef\blacksquare="0\msx@04
\mathchardef\centerdot="2\msx@05
\mathchardef\lozenge="0\msx@06
\mathchardef\blacklozenge="0\msx@07
\mathchardef\circlearrowright="3\msx@08
\mathchardef\circlearrowleft="3\msx@09
\mathchardef\rightleftharpoons="3\msx@0A
\mathchardef\leftrightharpoons="3\msx@0B
\mathchardef\boxminus="2\msx@0C
\mathchardef\Vdash="3\msx@0D
\mathchardef\Vvdash="3\msx@0E
\mathchardef\vDash="3\msx@0F
\mathchardef\twoheadrightarrow="3\msx@10
\mathchardef\twoheadleftarrow="3\msx@11
\mathchardef\leftleftarrows="3\msx@12
\mathchardef\rightrightarrows="3\msx@13
\mathchardef\upuparrows="3\msx@14
\mathchardef\downdownarrows="3\msx@15
\mathchardef\upharpoonright="3\msx@16

\mathchardef\downharpoonright="3\msx@17
\mathchardef\upharpoonleft="3\msx@18
\mathchardef\downharpoonleft="3\msx@19
\mathchardef\rightarrowtail="3\msx@1A
\mathchardef\leftarrowtail="3\msx@1B
\mathchardef\leftrightarrows="3\msx@1C
\mathchardef\rightleftarrows="3\msx@1D
\mathchardef\Lsh="3\msx@1E
\mathchardef\Rsh="3\msx@1F
\mathchardef\rightsquigarrow="3\msx@20
\mathchardef\leftrightsquigarrow="3\msx@21
\mathchardef\looparrowleft="3\msx@22
\mathchardef\looparrowright="3\msx@23
\mathchardef\circeq="3\msx@24
\mathchardef\succsim="3\msx@25
\mathchardef\gtrsim="3\msx@26
\mathchardef\gtrapprox="3\msx@27
\mathchardef\multimap="3\msx@28
\mathchardef\therefore="3\msx@29
\mathchardef\because="3\msx@2A
\mathchardef\doteqdot="3\msx@2B

\mathchardef\triangleq="3\msx@2C
\mathchardef\precsim="3\msx@2D
\mathchardef\lesssim="3\msx@2E
\mathchardef\lessapprox="3\msx@2F
\mathchardef\eqslantless="3\msx@30
\mathchardef\eqslantgtr="3\msx@31
\mathchardef\curlyeqprec="3\msx@32
\mathchardef\curlyeqsucc="3\msx@33
\mathchardef\preccurlyeq="3\msx@34
\mathchardef\leqq="3\msx@35
\mathchardef\leqslant="3\msx@36
\mathchardef\lessgtr="3\msx@37
\mathchardef\backprime="0\msx@38
\mathchardef\risingdotseq="3\msx@3A
\mathchardef\fallingdotseq="3\msx@3B
\mathchardef\succcurlyeq="3\msx@3C
\mathchardef\geqq="3\msx@3D
\mathchardef\geqslant="3\msx@3E
\mathchardef\gtrless="3\msx@3F
\mathchardef\sqsubset="3\msx@40
\mathchardef\sqsupset="3\msx@41
\mathchardef\trianglerighteq="3\msx@44
\mathchardef\trianglelefteq="3\msx@45
\mathchardef\bigstar="0\msx@46
\mathchardef\between="3\msx@47
\mathchardef\blacktriangledown="0\msx@48
\mathchardef\blacktriangleright="3\msx@49
\mathchardef\blacktriangleleft="3\msx@4A
\mathchardef\blacktriangle="0\msx@4E
\mathchardef\triangledown="0\msx@4F
\mathchardef\eqcirc="3\msx@50
\mathchardef\lesseqgtr="3\msx@51
\mathchardef\gtreqless="3\msx@52
\mathchardef\lesseqqgtr="3\msx@53
\mathchardef\gtreqqless="3\msx@54
\mathchardef\Rrightarrow="3\msx@56
\mathchardef\Lleftarrow="3\msx@57
\mathchardef\veebar="2\msx@59
\mathchardef\barwedge="2\msx@5A
\mathchardef\doublebarwedge="2\msx@5B
\mathchardef\angle="0\msx@5C
\mathchardef\measuredangle="0\msx@5D
\mathchardef\sphericalangle="0\msx@5E
\mathchardef\varpropto="3\msx@5F
\mathchardef\smallsmile="3\msx@60
\mathchardef\smallfrown="3\msx@61
\mathchardef\Subset="3\msx@62
\mathchardef\Supset="3\msx@63
\mathchardef\Cup="2\msx@64

\mathchardef\Cap="2\msx@65

\mathchardef\curlywedge="2\msx@66
\mathchardef\curlyvee="2\msx@67
\mathchardef\leftthreetimes="2\msx@68
\mathchardef\rightthreetimes="2\msx@69
\mathchardef\subseteqq="3\msx@6A
\mathchardef\supseteqq="3\msx@6B
\mathchardef\bumpeq="3\msx@6C
\mathchardef\Bumpeq="3\msx@6D
\mathchardef\lll="3\msx@6E

\mathchardef\ggg="3\msx@6F

\mathchardef\circledS="0\msx@73
\mathchardef\pitchfork="3\msx@74
\mathchardef\dotplus="2\msx@75
\mathchardef\backsim="3\msx@76
\mathchardef\backsimeq="3\msx@77
\mathchardef\complement="0\msx@7B
\mathchardef\intercal="2\msx@7C
\mathchardef\circledcirc="2\msx@7D
\mathchardef\circledast="2\msx@7E
\mathchardef\circleddash="2\msx@7F
\def\ulcorner{\delimiter"4\msx@70\msx@70 }
\def\urcorner{\delimiter"5\msx@71\msx@71 }
\def\llcorner{\delimiter"4\msx@78\msx@78 }
\def\lrcorner{\delimiter"5\msx@79\msx@79 }
\def\yen{\mathhexbox\msx@55 }
\def\checkmark{\mathhexbox\msx@58 }
\def\circledR{\mathhexbox\msx@72 }
\def\maltese{\mathhexbox\msx@7A }
\mathchardef\lvertneqq="3\msy@00
\mathchardef\gvertneqq="3\msy@01
\mathchardef\nleq="3\msy@02
\mathchardef\ngeq="3\msy@03
\mathchardef\nless="3\msy@04
\mathchardef\ngtr="3\msy@05
\mathchardef\nprec="3\msy@06
\mathchardef\nsucc="3\msy@07
\mathchardef\lneqq="3\msy@08
\mathchardef\gneqq="3\msy@09
\mathchardef\nleqslant="3\msy@0A
\mathchardef\ngeqslant="3\msy@0B
\mathchardef\lneq="3\msy@0C
\mathchardef\gneq="3\msy@0D
\mathchardef\npreceq="3\msy@0E
\mathchardef\nsucceq="3\msy@0F
\mathchardef\precnsim="3\msy@10
\mathchardef\succnsim="3\msy@11
\mathchardef\lnsim="3\msy@12
\mathchardef\gnsim="3\msy@13
\mathchardef\nleqq="3\msy@14
\mathchardef\ngeqq="3\msy@15
\mathchardef\precneqq="3\msy@16
\mathchardef\succneqq="3\msy@17
\mathchardef\precnapprox="3\msy@18
\mathchardef\succnapprox="3\msy@19
\mathchardef\lnapprox="3\msy@1A
\mathchardef\gnapprox="3\msy@1B
\mathchardef\nsim="3\msy@1C
\mathchardef\napprox="3\msy@1D
\mathchardef\nsubseteqq="3\msy@22
\mathchardef\nsupseteqq="3\msy@23
\mathchardef\subsetneqq="3\msy@24
\mathchardef\supsetneqq="3\msy@25
\mathchardef\subsetneq="3\msy@28
\mathchardef\supsetneq="3\msy@29
\mathchardef\nsubseteq="3\msy@2A
\mathchardef\nsupseteq="3\msy@2B
\mathchardef\nparallel="3\msy@2C
\mathchardef\nmid="3\msy@2D
\mathchardef\nshortmid="3\msy@2E
\mathchardef\nshortparallel="3\msy@2F
\mathchardef\nvdash="3\msy@30
\mathchardef\nVdash="3\msy@31
\mathchardef\nvDash="3\msy@32
\mathchardef\nVDash="3\msy@33
\mathchardef\ntrianglerighteq="3\msy@34
\mathchardef\ntrianglelefteq="3\msy@35
\mathchardef\ntriangleleft="3\msy@36
\mathchardef\ntriangleright="3\msy@37
\mathchardef\nleftarrow="3\msy@38
\mathchardef\nrightarrow="3\msy@39
\mathchardef\nLeftarrow="3\msy@3A
\mathchardef\nRightarrow="3\msy@3B
\mathchardef\nLeftrightarrow="3\msy@3C
\mathchardef\nleftrightarrow="3\msy@3D
\mathchardef\divideontimes="2\msy@3E
\mathchardef\varnothing="0\msy@3F
\mathchardef\nexists="0\msy@40
\mathchardef\mho="0\msy@66
\mathchardef\thorn="0\msy@67
\mathchardef\beth="0\msy@69
\mathchardef\gimel="0\msy@6A
\mathchardef\daleth="0\msy@6B
\mathchardef\lessdot="3\msy@6C
\mathchardef\gtrdot="3\msy@6D
\mathchardef\ltimes="2\msy@6E
\mathchardef\rtimes="2\msy@6F
\mathchardef\shortmid="3\msy@70
\mathchardef\shortparallel="3\msy@71
\mathchardef\smallsetminus="2\msy@72
\mathchardef\thicksim="3\msy@73
\mathchardef\thickapprox="3\msy@74
\mathchardef\approxeq="3\msy@75
\mathchardef\succapprox="3\msy@76
\mathchardef\precapprox="3\msy@77
\mathchardef\curvearrowleft="3\msy@78
\mathchardef\curvearrowright="3\msy@79
\mathchardef\digamma="0\msy@7A
\mathchardef\varkappa="0\msy@7B
\mathchardef\hslash="0\msy@7D
\mathchardef\hbar="0\msy@7E
\mathchardef\backepsilon="3\msy@7F
\def\Bbb{\ifmmode\let\next\Bbb@\else
 \def\next{\errmessage{Use \string\Bbb\space only in math mode}}\fi\next}
\def\Bbb@#1{{\Bbb@@{#1}}}
\def\Bbb@@#1{\fam\msyfam#1}

\def\inv{^{\raise.15ex\hbox{${
  \scriptscriptstyle -}$}\kern-.05em 1}}

\def\Dsl{\,\raise.15ex\hbox{$/$}\mkern-13.5mu D}
\def\dsl{\raise.15ex\hbox{$/$}\kern-.57em\hbox{$\partial$}}

\def\lspace{\ifx\answ\bigans{}\else\qquad\fi}
\def\del{\partial}

 \def\CH{\hbox{{$\cal H$}}}
 
 \def\CS{\hbox{{$\cal S$}}}

 \def\CD{\hbox{{$\cal D$}}}

\def\lform{\hbox{$\sqcup$}\llap{\hbox{$\sqcap$}}}
\def\darr#1{\raise1.5ex\hbox{$\leftrightarrow$}
\mkern-16.5mu #1}

\def\h{{{1\over2}}}

\def\INT{{\textstyle \int\kern-.642em\int}}

\def\R{{\Bbb R}}
\def\C{{\Bbb C}}
\def\Z{{\Bbb Z}}

\def\eps{{\epsilon}}

\def\cocross{{>\!\!\!\triangleleft}}

\def\tens{\mathop{\otimes}}

\def\la{{\triangleright}}

\def\isom{{\cong}}

\def\id{{\rm id}}

\def\Lin{{\rm Lin}}

\def\nquad{{\!\!\!\!\!\!}}
\def\nqquad{\nquad\nquad}
\def\eqn#1#2{\begin{equation}#2\label{#1}\end{equation}}

\def\o{{}_{(1)}}\def\t{{}_{(2)}}\def\th{{}_{(3)}}

\def\new#1{\goodbreak\goodbreak\bigskip
\noindent{\bf #1}}

\def\text#1{\mbox{\rm #1}}
\def\note#1{}

\def\blacksquare{{\lform}}
\def\frac#1#2{{{#1\over#2}}}

\def\proof{\goodbreak\noindent{\bf Proof\quad}}
\def\endproof{{\ $\lform$}\bigskip }

\def\align#1{\begin{eqnarray*}#1\end{eqnarray*}}

\def\alignn#1#2{\begin{eqnarray}\label{#1}#2
\end{eqnarray}}

\makeatother

\documentstyle[11pt]{article}
\newtheorem{lemma}{Lemma}[section]
\newtheorem{propos}[lemma]{Proposition}

\newtheorem{theorem}[lemma]{Theorem}

\newtheorem{corol}[lemma]{Corollary}
\newtheorem{defin}[lemma]{Definition}

\begin{document}\baselineskip 24pt

{\ }\hskip 4.7in DAMTP/92-20 
\vspace{.5in}

\begin{center} {\bf QUANTUM RANDOM WALKS AND TIME-REVERSAL}
\baselineskip 13pt{\ }\\
{\ }\\ S. Majid\footnote{SERC Fellow and Drapers Fellow of Pembroke College,
Cambridge}\\ {\ }\\
Department of Applied Mathematics\\
\& Theoretical Physics\\ University of Cambridge\\ Cambridge CB3 9EW, U.K.
\end{center}

\begin{center}
March 1992\end{center}
\vspace{10pt}
\begin{quote}\baselineskip 13pt
\noindent{ABSTRACT} Classical random walks and Markov processes are easily
described by Hopf algebras. It is also known that groups and Hopf algebras
(quantum groups) lead to classical and quantum diffusions. We study here the
more primitive notion of a quantum random walk associated to a general Hopf
algebra and show that it has a simple physical interpretation in quantum
mechanics. This is by means of a representation theorem motivated from the
theory of Kac algebras:
If $H$ is any Hopf algebra, it may be realised in $\Lin(H)$ in such a way that
$\Delta h=W(h\tens 1)W^{-1}$ for an operator $W$. This $W$ is interpreted as
the time evolution operator for the system at time $t$ coupled
quantum-mechanically to the system at time $t+\delta$. Finally, for every Hopf
algebra there is a dual one, leading us to a duality operation for quantum
random walks and quantum diffusions and a notion of the coentropy of an
observable. The dual system has its time reversed with respect to the original
system, leading us to a CTP-type theorem.
\end{quote}
\baselineskip 24.5pt

\section{Introduction}

A Hopf algebra is an algebra $H$ over a field $k$, equipped with algebra
homomorphisms $\Delta:H\to H\tens H$ (the comultiplication) and $\eps:H\to k$
(the counit) such that
\eqn{comult}{(\Delta\tens\id)\Delta=(\id\tens\Delta)\Delta,\qquad
(\eps\tens\id)\Delta=\id=(\id\tens\eps)\Delta.}
In addition, there is a linear map $S:H\to H$ (the antipode) obeying
$\cdot(S\tens\id)\Delta=1\eps=\cdot(\id\tens S)$.
The axioms (\ref{comult}) are obtained by writing out the axioms of an algebra
and reversing the arrows. Thus the algebra multiplication $\cdot:H\tens H\to H$
is supplemented by a `reversed-multiplication' $\Delta$. This means that the
algebra axioms are supplemented in such a way as to  restore some kind of
time-reversal or input-output symmetry. Hopf algebras are perhaps among the
simplest systems where this kind of input-output symmetry can be formulated.

The goal of this paper is to use Hopf algebras to explore this general notion
of input-output symmetry in a precise setting, namely that of random walks and
Markov processes. Thus our algebras $H$ will always be thought of as algebras
of observables (i.e. as if obtained from boolean algebras or from probability
spaces, or quantum algebras of observables in the non-commutative case). In
this context, asking that the algebra is a Hopf algebra really does restore an
input-output symmetry to the system. An element $h\in H$ is viewed as a random
variable and $\Delta h\in H\tens H$ expresses a linear combination of composite
random variables whose joint evaluation would lead to an evaluation of $h$.
Thus, while
algebras are traditionally used in formulations of (intuitionistic) deductive
logic, with the multiplication expressing necessity of a conclusion, the
presence of a Hopf algebra puts us into a framework of modal logic, with the
comultiplication expressing (in some sense) the notion of `possibility' to
supplement the notion of `necessity'.
We will examine these issues in a probabilistic setting rather than a logical
one, based on classical and quantum random processes.

The idea of thinking about Hopf algebras in this context is not a new one. At
least in the context of stochastic calculus, it is well known that Hopf
algebras provide a tool for formulating random walks and generalizing them.
Indeed, this was one of the motivations behind the development of Hopf algebras
some years ago, and in more recent times has lead to quantum diffusions and
other processes based on Hopf
algebras\cite{ASW:ind}\cite{Sch:whi}\cite{Bia:som}. From
a mathematical point of view there is no particular obstruction to going ahead
and formulating the same constructions when the Hopf algebra is an arbitrary
(possibly non-commutative) one. In particular, random walks make sense even
when the Hopf algebra is non-commutative. On the other hand, it is not at all
clear {\em a priori} if such generalized random walks are anything more than
mathematical deformations, i.e. if they are physical quantum processes. It is
this question that we address in the present paper.

Our main result in this direction is in Section~3 and is an algebraic operator
realization theorem for arbitrary Hopf algebras. Some of the ideas behind the
realization theorem come from the theory of Kac algebras,
see\cite{EnoSch:dua}\cite{BaaSka:uni}\cite{DaeKee:Yan}, but a feature of our
new treatment is that neither the $*$-structure nor a Hilbert space need be
built in from the start. Thus, our construction should be of interest to Hopf
algebraists and in the theory
of discrete random processes, as well as in the quantum context. In a quantum
mechanical context  it means that every Hopf algebra $H$ leads in a reasonably
convincing way to a quantum random walk based on a quantum evolution operator
$W$ (which we build from $H$). This is such that a step of the random walk on
$H$, considered actively as an operation on $H$,
consists of the following. First, we embed $h\in H$ as $h\tens 1\in H\tens H$.
Here the first copy of $H$ is the algebra of observables at time $t$ and the
second copy is the algebra of observables at time $t+\delta$ (i.e. one step
further in time). Then we evolve $h\tens 1$ by the quantum evolution operator
$W$ of the joint system $H\tens H$. Finally, we take an expectation value in
the first copy of $H$ to leave us in the second copy at $t+\delta$. This is the
quantum step of a quantum random walk. Our theorem asserts that a random walk
on any Hopf algebra can always be put into this form.

Section~4 then proceeds to study the impact of Hopf algebra duality in this
context. Every Hopf algebra has a dual one, and hence every such quantum random
walk has a dual one. We find that it is truly time-reversed with respect to the
original. The ideas here are based on the notion of observable-state symmetry
(quantum `Mach principle') developed in the quantum-gravitational context in
\cite{Ma:non}\cite{Ma:pla}\cite{Ma:pri}.

We begin in Section~2 with a brief introduction to how Hopf algebra methods can
be used to do classical random walks. Nothing in Section~2.1 should be new to
experts, however we have not been able to find a suitable treatment elsewhere.
We show how to use the Hopf algebra of the real line to easily derive the
diffusion equation for Brownian motion from the limit of a random walk given by
stepping to the left or right with probability $p,(1-p)$. Section~2.2 shows how
straightforward $q$-deformation techniques lead immediately to a $q$-deformed
random walk on $\R$ and a limiting $q$-Brownian process. The latter result can
be compared with the quantum stochastic process related to the Azema martingale
and based on a similar (but different) Hopf algebra in \cite{Sch:whi}. Our more
primitive treatment as the limit of a discrete walk seems to be new.

Let us note that the question of when a (quantum) algebra of observables is a
Hopf algebra has already been studied by the author in a very different context
through a series of papers
\cite{Ma:non}\cite{Ma:phy}\cite{Ma:hop}\cite{Ma:pla}. We studied particles
moving on curved spaces and found that not every background metric admits this
possibility of a Hopf algebra structure for the quantization of a particle in
it. For example, for a differentiable 1+1 dimensional
quantum dynamical system we looked for Hopf algebras of self-dual type and
found that the only possibility was for the background to be that of a
black-hole type metric\cite{Ma:pla}. Mathematically, the existence of a
non-cocommutative Hopf algebra structure on the (non-commutative) quantum
algebra of observables, corresponds to a non-Abelian group-structure on
`phase-space'. Here we regard the quantum algebra of observables in the fashion
of non-commutative geometry as if it is the ring of functions on some space.
The phase space does not exist
in an ordinary sense for in this case the functions on it would be commutative.
Physically, the reason is that the position and co-ordinate functions in phase
space can no longer be determined simultaneously because of Heisenberg's
uncertainty principle. Thus, the Hopf algebra structure corresponds to a
non-Abelian group structure on phase space but in a language general enough to
allow the space to be a quantum space rather than a classical one. The
non-Abelianess
means that the phase-space as a (quantum) space is curved, and hence led to
models exhibiting a unification of quantum effects with gravitational ones.
These models in \cite{Ma:non}\cite{Ma:pla} also exhibited a remarkable duality
phenomenon, based on Hopf algebra duality and interchanging microscopic
(quantum) physics with macroscopic (gravitational physics)\cite{Ma:pri}. As a
byproduct, we were led to a large class of non-commutative and
non-cocommutative Hopf algebras which were quite different from the more
popular quasitriangular Hopf algebras (quantum groups) of Drinfeld and Jimbo
etc. related to Yang-Baxter equations and such topics.

Thus, it is the same mathematical structure, namely an algebra of observables
that is a Hopf algebra, which has these two interpretations, (i) in terms of
gravitational physics as in the authors previous work and (ii) in terms of
probability theory as here. Either application alone would lead us to study
Hopf algebras and the fact that the same structure serves both indicates the
possibility of a remarkable unification of the two topics. Roughly speaking,
generalized (quantum) Riemannian geometry can be reformulated in terms of the
language of random walks, and vice-versa. This is one of the long-term
motivations behind the present work, and will be further developed elsewhere.
Here we concentrate only on the
probabilistic interpretation. Both (i) and (ii) are quite different again from
the standard interpretation of quasitriangular Hopf algebras in the theory of
inverse scattering - where the quantum group is not at all a quantum algebra of
observables but rather a quantum symmetry. This role of quantum symmetry
represents a third potential unification made possible by Hopf algebras, but
not something that we shall develop here. On the other hand, there is a
tangential connection between related quantum spin chains and non-commutative
Markov processes which can be mentioned.

\new{Acknowledgments} I thank K.R. Parthasarathy and several others for
valuable discussions during the December 1991 meeting in Oberwolfach.

\section{Random Walks and Markov Processes via Hopf Algebras}

In this preliminary section we explain in elementary terms how Hopf algebras
can be used to describe random walks and Markov processes. Even in simple and
well-known cases they encode the required computation in a very direct way. The
reason for this is quite fundamental as explained in the introduction.

\subsection{Brownian Motion}

We begin by recalling some standard ideas from classical probability. To be
concrete we focus on $\R$ (the real line) as our probability space. A
probability density function for us is simply a pointwise positive function
$\rho\in L^1(\R)$ such that $\int dx \rho(x)=1$. We also allow atomic measures
($\delta$-functions). We can use such a $\rho$ to define a random walk as
follows: at $x_0=0$, $x_{i+1}=x_i+X$ where $X$ is chosen randomly with
probability distribution $\rho$. The routine question to ask is: what is the
probability distribution after $n$ steps, i.e. of $x_n$? Clearly it is given by
a probability density $\rho^{n}$ (say) defined by
\eqn{e1}{ \rho^{n}(x)=\int_{y_1+\cdots y_n=x} dy_1\cdots dy_n \rho(y_1)\cdots
\rho(y_n).}
Note that in asking after the system at various steps, we are always asking the
same mathematical question, namely the distribution of a random variable on the
real line. Thus we take the point of view that the question, which we denote
$X$, is not changing with each step. Rather, it is the probability distribution
of $X$ which is changing with each
step. This $X$ is the abstract position observable of the random walk, i.e.
`where the particle is at' (statistically speaking). Its expectation value and
moments etc are changing with each time step. The distribution (\ref{e1})
immediately gives
\eqn{e2}{ <1>_{\rho^n}=<1>_{\rho}^n=1,\quad <X>_{\rho^n}=n<X>_\rho,\quad
<X^2>_{\rho^n}=n(<X^2>_\rho-<X>_\rho^2)+n^2<X>^2_\rho}
where $<\ >$ denotes expectation value in the marked state. From this we see
that the mean position is increasing with each step, as would be expected.
Also, if we define $Y^{(n)}={1\over n}X$ then $<Y^{(n)}>_{\rho^n}=<X>_{\rho}$
is independent of $n$ while its variance
$<Y^{(n)}{}^2>_{\rho^n}-<Y^{(n)}>_{\rho^n}^2={1\over n}(<X^2>_\rho-<X>_\rho^2)$
decreases. Similarly the higher variances decrease, so that the rescaled
variables $Y^{(n)}$ become more and more sharply peaked. This is the basis of
the central limit theorem.

This elementary computation can be done by means of Hopf algebras as follows.
Firstly, we regard the abstract question $X$ (roughly speaking) as an element
of $L^{\infty}(\R)$ (the Hopf-von Neumann algebra of functions on the real
line). It is defined as the coordinate function $X(x)=x$. Now, a probability
distribution assigns to any power $X^n$, or more precisely any bounded function
$f(X)$ an expectation value $<f(X)>_\rho=\int dx \rho(x) f(x)$. This gives a
positive linear functional
$\phi$ on $L^\infty(\R)$, corresponding to $\rho$ via $\phi(f)=<f(X)>_\rho$.
Positive means $\phi(f^*f)>0$ for all $f\ne 0$ and corresponds to $\rho$ real
and positive. The normalization corresponds to $\phi(1)=1$. Thus we can work
directly with the Hopf algebra $L^\infty(\R)$ in place of random variables, and
$\phi$ a (normal, unital) state on $L^\infty(\R)$ in place of $\rho$. This is a
familiar point of view in the context of classical mechanics (and quantum
mechanics) and $L^\infty(\R)$ here is the algebra of observables of the system.

The Hopf algebra structure on $L^\infty(\R)$ is $(\Delta f)(x,y)=f(x+y)$, where
$L^\infty(\R)\tens L^\infty(\R)=L^\infty(\R\times\R)$. It expresses the group
law on $\R$ in algebraic terms. The co-ordinate function, for example, has the
additive comultiplication $(\Delta X)(x,y)=X(x+y)=x+y=(X\tens 1+1\tens X)(x,y)$
or as elements of the Hopf algebra,
\eqn{e3}{\Delta X=X\tens 1+1\tens X.}
The group law played a key role in (\ref{e1}) and likewise, $\Delta$ plays the
corresponding role in defining a new state $\phi^n$. Indeed, the
comultiplication on any Hopf algebra defines a multiplication in its dual, and
it is this multiplication (the convolution product) that we need. Explicitly,
\eqn{e4}{(\phi\psi)(f)=(\phi\tens\psi)(\Delta f), \quad {\em e.g.}\quad
\phi^n(f)=(\phi^{\tens n})(\Delta^{n-1}f).}
where $\Delta^{n-1}$ denotes $\Delta$ applied $n-1$ times. We will come to the
formal definition of a Hopf $*$-algebra in Section~2.3, but in our example (and
in general) the properties of the $*$-structure are such as to ensure that if
$\phi$ is positive then $\phi^n$ is also positive etc, i.e. also a state.
Associativity of this convolution algebra is ensured by the first of
(\ref{comult}), while $\eps$ in (\ref{comult}) is the identity state in the
algebra.

In these terms, the computation of (\ref{e2}) looks like
\align{ <1>_{\rho^n}&&=\phi^n(1)=(\phi^{\tens
n})\Delta^{n-1}1=(\phi\tens\cdots\tens\phi)(1\tens\cdots\tens 1)=\phi(1)^n=1\\
<X>_{\rho^n}&&=(\phi^{\tens
n})\Delta^{n-1}X=(\phi\tens\cdots\tens\phi)(X\tens\cdots\tens 1+\cdots+1\tens
\cdots \tens X)=n\phi(X)=n<X>_\rho\\
 <X^m>_{\rho^n}&&=(\phi^{\tens
n})\Delta^{n-1}X^m=(\phi\tens\cdots\tens\phi)(\sum_{i_1+\cdots+i_n=m} {m\choose
i_1\cdots i_n}X^{i_1}\tens\cdots\tens X^{i_n})\\
&&=\sum {m\choose i_1\cdots i_n}\phi(X^{i_1})\cdots\phi(X^{i_n})=\sum {m\choose
i_1\cdots i_n}<X^{i_1}>_\rho\cdots<X^{i_n}>_\rho}
Thus we reproduce the results above in this new language. The computation in
this form uses nothing other than the two equations (\ref{e3}) and (\ref{e4}).
Not only does it bypass unpleasant convolution integrals as in (\ref{e1}), but
it is conceptually rather cleaner also. The conceptual picture is that
expectation value of $X$ after $n$ steps is simply the expectation value of
$\Delta^{n-1} X$ in the tensor product system $L^\infty(\R)^{\tens
n},\phi^{\tens n}$. The tensor product system is the system for $n$ independent
random variables, each with the same distribution $\phi$, and the iterated
comultiplication embeds the algebra of observables for our one particle into
this system. If we denote these co-ordinate functions that generate our
$n$-fold tensor product by
\eqn{e5}{X_i=1\tens\cdots \tens X\tens \cdots\tens 1}
(embedded in the $i$'th position), then the random variable $X$ of our one
particle embeds as $X_1+X_2+\cdots+X_n$. Regarding $\Delta^{n-1}$ as
understood, we can simply write $X=X_1+\cdots+X_n$. Here the $X_i$ are $n$
independent random variables.

To demonstrate the power of this formalism further, let us compute a specific
example. In terms of $\rho$ we take an atomic measure with
$\rho=p\delta_a+(1-p)\delta_{-a}$, i.e. peaked as $\delta$-functions at $a,-a$.
In terms of $\phi$ we have
\eqn{e6}{\phi(f)=(p\phi_a+(1-p)\phi_{-a})(f)=pf(a)+(1-p)f(-a)}
where $\phi_a(f)=f(a)$ is the linear map on $L^\infty(\R)$ given by evaluation
at $a$. We can also introduce the linear map $D(f)=f'(0)$. It is not a state
but we can still view it as a (densely defined) map on $L^\infty(\R)$ and make
the convolution product etc as above, i.e. we formally view it in the
convolution algebra. Then the form of $\Delta$ in (\ref{e4}) gives
\eqn{e7}{D^n(f)=(D\tens\cdots\tens D)(\Delta^{n-1}f)={\del\over\del
x_1}|_0\cdots {\del\over\del x_n}|_0 f(x_1+\cdots+x_n)=f^{(n)}(0).}
Hence in this convolution algebra we have the expansion
\eqn{e8}{\phi_a=\eps+aD+{a^2\over 2!}D^2+\cdots}
(Taylors theorem). We are now ready to compute the system after $n$ steps (and
its limit) as described by the states
\eqn{e9}{\phi^n=(\eps+2a(p-\h)D+{a^2\over 2!}D^2+\cdots)^n=(\eps+{ct\over
n}D+{t\alpha\over n}D^2+\cdots)^n\to \phi^\infty=e^{t\alpha D^2+{ct}D}}
where $t,c,\alpha$ are defined by $t=n\delta$, ${a^2\over 2}={t\alpha\over
n}=\alpha\delta$ and $2a(p-\h)= {tc\over n}=\delta c$ and we send $a\to 0$ and
$n\to\infty$ (or $\delta\to 0$) with $t,c,\alpha$ fixed.

This limiting state describes our random walk after an infinite number of
steps, with the step viewed as a step in time of size $\delta$, which tends to
zero, i.e. it is the continuous limit of the random walk. We can still evaluate
our observables in this limit,
\eqn{e10}{\phi^\infty(f)=(e^{t\alpha D^2+{ct}D})(f)=(e^{t\alpha {d\over
dx}^2+{ct}{d\over dx}}f)(0)}
and we can also compute the distribution $\rho^\infty$ corresponding to
$\phi^\infty$. Formally, if $\phi$ is any state we can formally get back the
corresponding density by $\rho(x)=\phi(\delta_x)$ (the $\delta$-function at
$x$, which of course has to be approximated to lie in $L^\infty(\R)$). We have
\eqn{e11}{\rho^\infty(y)=\phi^\infty(\delta_y)=(e^{t\alpha {d\over
dx}^2+{ct}{d\over dx}}\delta_y)\vert_{x=0}=(4\pi\alpha
t)^{-\h}e^{-{(y-ct)^2\over 4\alpha t}}}
where the right hand side is the unique solution $G(y,t)$ of the diffusion
equation ${\del\over\del t}G(y,t)=\alpha {\del^2\over\del
y^2}G(y,t)-{c}{\del\over\del y}G(y,t)$. Another way to see that this must
coincide with $\rho^\infty$ is that the latter is characterized by $\int
\rho^\infty(x)f(x)=e^{t\alpha D^2+{ct}D}(f)$ for all $f$: Differentiating this
$\del\over\del t$ we have
\alignn{e12}{ \int ({\del\over\del t}\rho^\infty)(x)f(x)dx&&=((\alpha D^2+{
c}D)e^{t\alpha D^2+{ct}D})(f)\nonumber\\
&&=\int \rho^\infty(x)(\alpha {d^2\over dx^2}+ {c}{d\over dx})
f(x)dx\nonumber\\
&&=\int ((\alpha {d^2\over dx^2}-{c}{d\over dx})\rho^\infty (x))f(x)dx}
Where $(D\tens\id)\Delta f(X)=f'(X)$. Our derivation of (\ref{e11}) can be
compared with more standard derivations based on Stirling's formula etc. In our
case, we used only a limit of the form $(1+{Z\over n})^n\to e^Z$ in deriving
(\ref{e9}).

These elementary computations demonstrate the usefulness and directness of this
approach based on Hopf algebras. A remarkable fact can now be observed: in
these computations we do not need to assume that the Hopf algebra is
commutative! Thus, if $H$ is any Hopf algebra, the convolution product
(\ref{e4}) for elements $\phi,\psi$ of $H^*$ make perfectly good sense. We can
also take continuum limits by expanding $\phi$ and using the same limiting
procedure as in (\ref{e9})
without assuming commutativity. These remarks fit well into the overall context
of `non-commutative' or `quantum' probability,  where we replace
$L^\infty(\Omega)$  for a classical probability space $\Omega$ by a
non-commutative von Neumann algebra: we can do random walks by working with
not-necessarily commutative Hopf-von Neumann algebras. Such generalized random
walks can be called `non-commutative' or `quantum' random walks.

This situation can be viewed as some of the underlying motivation behind, for
example, the notion of a quantum stochastic process. Examples of the latter are
indeed known in connection with Hopf algebras\cite{Sch:whi}, though how they
might be obtained as the limit of random walks seems to be less well studied,
but see for example \cite{LinPar:pas}. Moreover, whether such generalized
non-commutative random walks can be viewed as actual quantum mechanical
processes is also an open question. These are two of the issues to be addressed
in Sections~3,4.

\subsection{$q$-Brownian Motion}

We now give a non-commutative example. The above random walk leads to Brownian
motion (given by the diffusion equation): the non-commutative example will
constitute a $q$-deformation of it, and can be called $q$-Brownian motion. A
related example has been studied as a quantum stochastic process in
\cite{Sch:whi}\cite{Par:aze} under the heading of the Azema process. Our more
primitive treatment as the limit of a random walk seems to be new and in any
case is
based on a different Hopf algebra. Note that a $*$-structure was implicit in
the above (the states are taken positive) and this continues to be so in our
setting now even though the algebra is non-commutative. As in quantum
mechanics, the observables of most interest are the self-adjoint ones and
states should be positive in the sense $\phi(h^*h)\ge 0$ for all $h$ in the
algebra. We assume they are normalised as $\phi(1)=1$.

For our Hopf algebra $H$ we take generators $X,g,g^{-1}$ and relations,
comultiplication, counit, antipode and $*$-structure
\eqn{e13}{ gX=qXg,\quad gg^{-1}=1=g^{-1}g,\quad \Delta g=g\tens g,\quad \Delta
X=X\tens g^{-1}+g\tens X,\quad \eps g=1,\ \eps X=0.}
\eqn{e13.5}{SX=-q^{-1}X,\qquad Sg=g^{-1},\qquad X^*=X,\quad g^*=g,\quad
q^*=q^{-1}.}
 This is a standard Hopf algebra \cite{Taf} but note that $q$ is a parameter of
modulus 1 for the $*$-structure that we need. The elements $X^mg^i$ for $m\in
\Z_+,i\in \Z$ are a basis. The comultiplication should be compared with
(\ref{e3}). If $\phi$ is any state as above, the distribution $\phi^n$ after
$n$ steps is computed from (\ref{e4}) using the comultiplication $\Delta$. It
is given by embedding $X$ in $H^{\tens n}$ via $\Delta^{n-1}$ and applying
$\phi$ to each $H$. From (\ref{e13}) we have
\eqn{e14}{\Delta^{n-1}g=g\tens\cdots \tens g,\quad \Delta^{n-1}X=X\tens
g^{-1}\tens\cdots\tens g^{-1}+g\tens X\tens g^{-1}\tens\cdots\tens
g^{-1}+\cdots+g\tens\cdots g\tens X.}
If we write the right hand expression as $\Delta^{n-1}X=X_1+X_2\cdots+X_n$
where $X_1=X\tens g^{-1}\cdots\tens g^{-1}$, $X_2=g\tens X\tens
g^{-1}\cdots\tens g^{-1}$ etc, we see that the random variable describing the
position after $n$ steps is the sum of $n$ random variables $X_i$ embedded in
$H^{\tens n}$. They are not, however, independent in a usual sense. Instead,
\eqn{e15}{ X_iX_j=q^2X_jX_i,\qquad i>j}
from the relations in (\ref{e13}). This shows up when we look at higher
moments, where we need to compute $\Delta^{n-1}X^m$. This is a standard
computation in connection with the above Hopf algebra. We have
\eqn{e16}{\Delta^{n-1}X^m=\sum_{i_1+\cdots +i_n=m}{[m]_{q^2}!\over
[i_1]_{q^2}!\cdots [i_n]_{q^2}!} X_1^{i_1}\cdots X_n^{i_n},\quad
[i]_q={1-q^{n}\over 1-q}}
using the relations (\ref{e15}).

\begin{propos} The expectation values after $n$ steps are
\[ <g>_{\phi^n}=<g>_\phi^n,\quad
<X>_{\phi^n}=<X>_\phi{<g>^n_\phi-<g^{-1}>^n_\phi\over <g>_\phi-<g^{-1}>_\phi}\]
\[ <X^2>_{\phi^n}=<X^2>_\phi{<g^2>^n_\phi-<g^{-2}>^n_\phi\over
<g^2>_\phi-<g^{-2}>_\phi}+[2]_q<Xg>_\phi<g^{-1}X>_\phi{[n]_{<g^2>_\phi}-
[n]_{<g^{-2}>_\phi}\over <g^2>_\phi-<g^{-2}>_\phi}\]
These moments reduce to the usual results (\ref{e2}) for a random walk on $\R$
if we set $q\to 1$ and $g\to 1$ in a strong sense with $<g^{\pm 1}>_\phi\sim
<g>^{\pm 1}_\phi$ and  $<g^{\pm 2}>_\phi\sim <g>^{\pm 2}_\phi$.
\end{propos}
\proof The expectation of $g$ and $X$ follows at once from (\ref{e16}). For the
expectation of $X^2$ we have to write out  $\Delta X^2=\sum_{i=1}^n X_i^2+\sum
_{i<j}[2]_{q^2} X_iX_j$ from (\ref{e16}) explicitly in terms of the $X_i$. For
the first sum we have $X^2\tens g^{-2}\tens \cdots\tens g^{-2}+g^2\tens X\tens
g^{-2}\tens\cdots \tens g^{-2}+\cdots$ which gives the first term shown. For
the second term we note
\[X_iX_j=g^2\tens\cdots \tens g^2\tens Xg\tens 1\tens\cdots\tens 1\tens
g^{-1}X\tens g^{-2}\tens\cdots\tens g^{-1}\]
where $Xg$ is in the $i$'th position and $g^{-1}X$ is in the $j$'th. Applying
$\phi^{\tens n}$ gives a factor $<g^2>_\phi^{i-1}<g^{-2}>_\phi^{n-j}$. Summing
over $j$, we obtain for the second term in $<X^2>_{\phi^n}$ the result
$[2]_{q^2}<Xg>_\phi<g^{-1}X>_\phi$ times the expression
\align{&&\nqquad\sum_{i=1}^{n-1}<g^2>_\phi^{i-1}{1-<g^{-2}>_\phi^{n-i}\over
1-<g^{-2}>_\phi}\\
&&={1-<g^2>_\phi^{n-1}\over(1-<g^{-2}>_\phi)(1-<g^2>_\phi)}-{<g^{-2}>_\phi\over
(1-<g^{-2}>_\phi)}{(<g^2>^{n-1}_\phi-<g^{-2}>^{n-1}_\phi)\over
(<g^2>_\phi-<g^{-2}>_\phi)}\\
&&=(<g^2>_\phi-<g^{-2}>_\phi)^{-1}\left( {1-<g^2>^n_\phi\over
1-<g^2>_\phi}-{1-<g^{-2}>^n_\phi\over 1-<g^{-2}>_\phi}\right)}
after suitable reorganization of the partial fractions. Note that if we write
$<g^{\pm 2}>_\phi=1\pm2\delta+O(\delta)$ and take $\delta\to 0$ then this
expression tends to $n(n-1)\over 2$ as in Section~2.1
 \endproof

We can also compute the continuous limit of a random walk on this Hopf algebra,
as follows. For simplicity we assume  that $q$ is not a root of unity.
For our state $\phi$ we take
\eqn{e17}{ \phi (g^if(X)g^j)=pf(q^{i-j\over 2}a)+(1-p)f(-q^{i-j\over 2}a)}
for $a\in \R$ and $0\le p\le 1$. This reduces to the choice in Section~2.1 when
$i=j$. Note that $\phi((f(X)g^i)^*f(X)g^i)=\phi(g^i\bar f(X)f(X)g^i)=p\vert
f(a)\vert^2+(1-p)\vert f(-a)\vert^2$ so that $\phi$ is positive. Here, and
below, we can concentrate on $f$ given by polynomials or suitable power-series
in the $X$. As in Section~2.1, this algebraic approach is largely for
convenience: underlying it is a Hopf-von Neumann algebra similar to an
extension of $L^\infty(\R)$ above (this is needed to make sense of some of the
exponentials below).

In order to approximate $\phi$ we make use of a well known operator of
`$q$-differentiation' $\del_q$ cf\cite{And:ser} to define an operator
$\CD_q:H\to H$ by
\eqn{e17.5}{ \CD_q g^if(X)g^j=g^{i-\h}{f(qX)-f(q^{-1}X)\over
(q-q^{-1})X}g^{j-\h}q^{i-j\over 2}=g^{i-\h}(\del_q f)g^{j-\h}q^{i-j\over 2}}
where we have extended the action to $g$ in such a way that $\CD_q$ is a
$*$-preserving and (formally) a completely positive operator. Here the
$g^{-\h}$ is for notational convenience: it should be understood on one side or
the other as a factor $g^{-1}$ (in view of the relations (\ref{e13})). In
particular, we have
\eqn{e18}{  \CD_q g^iX^mg^j=[[m]]_q g^{i-1}X^{m-1}g^jq^{i-j\over 2}q^{m-1\over
2},\quad[[m]]_q={q^m- q^{-m}\over q-q^{-1}}.}
 Motivated by this we take a linear functional $D_q:H\to \C$ given by
evaluation of $\CD_q$ at $X=0,g=1$ and consider its iterated convolution
product according to (\ref{e4}). We have
\eqn{e18.5}{D_q (g^iX^mg^j)=\delta_{m,1}q^{i-j\over 2}}
\alignn{e19}{
D^n_q (g^iX^m g^j)&&=(D_q\tens\cdots D_q)\left((g^i\tens\cdots\tens
g^i)(\Delta^{n-1} X^m)(g^j\tens\cdots\tens g^j)\right)\nonumber\\
&&=\delta_{n,m}[m]_{q^2}! q^{i-(n-1+j)\over 2}q^{i-1-(n-2+j)\over 2}\cdots
q^{i-(n-1)-j\over 2}=\delta_{n,m}[[m]]_q! q^{{i-j\over 2}m}}
using (\ref{e14}) to obtain (\ref{e18.5}) and then (\ref{e16}),(\ref{e18.5}) to
obtain (\ref{e19}).

We also need a (well known) $q$-exponential $e_q$ defined with coefficients
$1/[[i]]_q!$ in place of the usual $1/i!$. From (\ref{e19}) we conclude in the
convolution algebra that
\eqn{e20}{(e_q^{aD_q} )(g^if(X)g^j)=f(q^{i-j\over 2}a).}

Hence we can approximate $\phi$ just as before, by
\eqn{e21}{\phi^n=(\eps+2a(p-\h)D_q+{a^2\over
[[2]]_q}D_q^2+\cdots)^n=(\eps+{ct\over n}D_q+{t\alpha\over
n}D_q^2+\cdots)^n,\quad \phi^\infty=e^{t\alpha D_q^2+{ct}D_q}}
where $t,c,\alpha$ are defined by $t=n\delta$, ${a^2\over
[[2]]_q}={t\alpha\over n}$ and $2a(p-\h)= {tc\over n}$ and we send $a\to 0$ and
$n\to\infty$ (or $\delta\to 0$) with $t,c,\alpha$ fixed as before.

We can then evaluate $\phi^\infty$ on elements of our algebra to compute
expectation values. We can also try to introduce a corresponding probability
density $\rho^\infty$ defined via
\eqn{e22}{\phi^\infty(f(X))=\int^\infty_{-\infty}\rho^\infty(x)f(x)dx,
\qquad\forall f}
with respect to ordinary integration (say). This means that we are visualizing
our deformed random walk with respect to the usual picture of $\R$, which is
embedded in $H$ as a subalgebra. This is not the only possibility (one could,
for example, use here a $q$-integration). Note also that
\eqn{e22.5}{\phi^\infty(g^if(X)g^i)=\phi^\infty(f(X))}
from (\ref{e19}), so (\ref{e22}) determines $\phi^\infty$ completely.
Proceeding with (\ref{e22}) and differentiating with respect to $t$ we have
\alignn{e23}{ \int ({\del\over\del t}\rho^\infty)(x)f(x)dx &&=((\alpha
D_q^2+{c}D_q)e^{t\alpha D_q^2+{ct}D_q})(f(X))\nonumber\\
&&=\phi^\infty((\alpha \CD_q^2+{c}\CD_q) f(X))\nonumber \\
&&=\int (\rho^\infty(x)\alpha \del_q^2+{c}\rho^\infty(x)\del_q) f(x)dx\nonumber
\\
&&=\int (\alpha \del^2_{q}\rho^\infty(x)-{ c}\del_{q}\rho^\infty (x))f(x)dx.}
where  we used
\eqn{e23.5}{(D_q\tens\id)\Delta =\CD_q.}
This is easily proven from (\ref{e14}) and (\ref{e18.5}). We also used
(\ref{e22.5}) and the identity
\alignn{e24}{&&\nquad\int h(x){f(qx)-f(q^{-1}x)\over (q-q^{-1})x}dx\nonumber\\
&&=\int {h(q^{-1}x')\over (q-q^{-1})x'}f(x')dx'-\int{h(qx')\over
(q-q^{-1})x'}f(x')dx'=-\int{h(qx)-h(q^{-1}x)\over (q-q^{-1})x}f(x)dx}
for all analytic functions $h,f$ such that the contours can be rotated for the
changes of variable (so that $\del_q^\dagger =-\del_{q}$ with respect to this
$L^2$ inner product on suitable test functions). Thus, $\rho^\infty$ is
characterized as the solution of
\eqn{e25}{{\del\over\del t} \rho^\infty(x,t)=\alpha
\del^2_{q}\rho^\infty(x,t)-{c}\del_{q} \rho^\infty(x,t).}
This is a $q$-deformed diffusion equation. Note that unlike the undeformed
case, however, our $\phi^\infty$ does not involve ordinary differentiation but
rather the non-local operator of $q$-differentiation, and hence there is no
reason to think that a smooth solution $\rho^\infty$ to (\ref{e25}) should
exist along familiar Gaussian lines. It can perhaps be interpreted
stochastically. Our elementary derivation of $q$-Brownian motion can be
compared with the treatment of the Azema martingale\cite{Par:aze} and its
quantum stochastic process in \cite{Sch:whi}.

\subsection{Transition Operators}

The last example should convince the reader that it is interesting to consider
random walks on non-commutative Hopf algebras, even if only as deformations of
standard ones. Note also that in the above random walks, we are not required to
use the same state $\phi$ for the distribution of each step of the walk. We
could just as easily have a collection of states
$\{\phi_1,\phi_2,\cdots,\phi_n\}$ to be used successively in each step of the
random walk. The distribution after $n$ steps is given by the convolution
product $\phi_1\phi_2\cdots\phi_n$, i.e. the tensor product
$\phi_1\tens\phi_2\cdots\tens\phi_n$ applied to the image via $\Delta^{n-1}$ in
$H^{\tens n}$. Such a random walk with distinct steps is said to be
non-stationary.

Closely related to random walks are Markov processes, and these too can be
built on Hopf algebras. In fact, for our purposes they are the same structure
from an equivalent `active' point of view. We briefly explain this now. Thus,
given a linear functional $\phi$ on $H$, we define $T_\phi:H\to H$ by
\eqn{MT}{T_\phi=(\phi\tens\id)\Delta}
This is equivalent to $\phi$, and also recovers the convolution multiplication
(\ref{e4}) as composition of operators
\eqn{Tmult}{\eps\circ T_\phi=\phi,\quad \phi\psi=\eps\circ T_\psi T_\phi}
on using the counity and coassociativity axioms (\ref{comult}) of
$\Delta,\eps$. Thus, our all-important convolution product is just composition
of a corresponding operator $T_\phi:H\to H$. It can be called the Markov
transition operator corresponding to $\phi$.

In terms of the Markov transition operator, the system evolves actively by
$T_\phi:H\to H$ and the final expectation values after $n$ steps are obtained
by applying $\eps$ to the evolved observables. Thus, if we have a random walk
with varying states $\phi_1,\cdots,\phi_n$ at each step, the corresponding
expectations after $n$ steps are
\eqn{Tmulti}{ <f>_{\phi_1\phi_2\cdots\phi_n}=\eps\circ T_{\phi_n}\cdots
T_{\phi_1}(f).}
An algebra $H$ equipped with a step-evolution operator, or operators $T:H\to H$
is a (possibly non-commutative) Markov process. We see that this is just an
`active' way of thinking about random walks, in which the observables rather
than the states evolve. We will say more about this in the next section.

In the above example, the role of the counit was in the evaluation at $X=0,g=1$
(see (\ref{e13})). Thus, $\CD_q=T_{D_q}$ is the operator ($q$-differentiation)
corresponding to the functional $D_q=\eps \circ\CD_q$ that was used, see
(\ref{e23.5}). Likewise, the transition operator $T_{\phi^\infty}$
corresponding to the limiting state is
\[ T_{\phi^\infty}=e^{t\alpha \CD_q^2+{ct}\CD_q},\]
i.e. the $q$-deformed time evolution.

This formalism of random walks and transition operators works for general Hopf
algebras over a field $k$. Finally, we formalize the situation regarding the
$*$-structure needed in a quantum mechanical setting. Recall that a $*$-algebra
means an algebra over $k=\C$ equipped with an antilinear involutive
anti-algebra homomorphism $*$. Then a Hopf $*$-algebra is a $*$-algebra $H$
such that
\eqn{hopfstar}{\Delta h^*=(\Delta h)^{*\tens *},\quad
\eps(h^*)=\overline{\eps(h)},\quad (S\circ *)^2=\id.}
Such $*$-structures on Hopf algebras have been emphasised by \cite{Wor:com},
among others. In this case (for $\phi$ positive), the transition operator
$T_\phi$ is a completely positive map \cite{ASW:ind}. Our example in
Section~2.2 fits into this setting and $\CD_q$ is completely positive, at least
formally.

Also, if $H$ is a Hopf algebra then $H^*$ (defined in a suitable way) is also a
Hopf algebra and the comultiplication in one is determined by the
multiplication in the other. This is just the origin of the convolution product
(\ref{e4}), but is also works the other way with $<\Delta \phi,h\tens
g>=<\phi,hg>$ and $<S\phi,h>=<\phi,Sh>$ with the comultiplication and antipode
of $H^*$ coming from $H$ (for this reason we write the evaluation symmetrically
as $<\ ,\ >$ and say that $H,H^*$ are dually paired Hopf algebras\cite[Sec.
1]{Ma:qua}). In the Hopf $*$-algebra case we have in addition
\eqn{dualstar}{ <\phi^*,h>=\overline{<\phi,(Sh)^*>}.}
It follows at once from this and (\ref{hopfstar}) that
\eqn{Mstar}{T_\phi(h)^*=T_{(S\phi)^*}(h^*)}
so that $T_\phi$ is $*$-preserving iff $\phi$ is {\em anti-self-adjoint} in the
sense
\eqn{ASA}{\phi^*=S\phi}
 in $H^*$. For our example of Section~2.2, the state $D_q$ is anti-self-adjoint
in this way so that $\CD_q$ was necessarily $*$-preserving, as easily seen by
direct computation. In summary, positivity of $\phi$ corresponds to complete
positivity of the corresponding transition operator, while
anti-self-adjointness corresponds to the transition operator being
$*$-preserving.

Let us note finally that Markov processes (like random walks and stochastic
processes) can be defined more generally than those based on Hopf algebras, for
example in the context of spin chains in \cite{Bia:som}. Also, their continuum
limit is related to a process of dilation of operators\cite{Kum:sur}.

\section{Operator Realization of Hopf Algebras}

 We have seen in the last section how any Hopf algebra $H$ equipped with a
linear functional $\phi$ can be interpreted formally as leading to a
generalized random walk or Markov process. In the random walk interpretation
the distribution of an observable $h\in H$ after $n$ steps is given by
embedding $h$ into $H^{\tens n}$ as $\Delta^{n-1}h$ and applying the
expectation value $\phi$ to each factor. The factors are the steps of the
random walk and $\Delta^{n-1}h$ is understood literally as the linear
combination of all  the ways to arrive at $h$ via the elements in $H^{\tens n}$
viewed as successive steps. In this section we explore this interpretation
further by proving a general result about the coproduct of an arbitrary Hopf
algebra.

Although we are primarily interested in the Hopf $*$-algebra situation needed
for quantum mechanics, our first result is more general and works over an
arbitrary field or commutative ring $k$. Perhaps this more general case will be
needed in discrete applications as well as in attempts to unify the present
considerations with Planck-scale physics as mentioned in the Introduction. If
we have to drop classical geometry at the Planck scale, then it is likely that
we will have to drop classical functional analysis too (based on $\R^n$) at
some point, and do everything algebraically. While the result is known in some
form at a Hopf $*$-algebra level\cite{BaaSka:uni}, our purely algebraic level
seems to be more novel.

\begin{theorem} Let $H$ be a Hopf algebra and view $H\subset \Lin(H)$ by
$h\mapsto h\la=L_h$ (the left regular representation $L_h(g)=hg$). In the
algebra $\Lin(H\tens H)\supseteq\Lin(H)\tens\Lin(H)$ there is an invertible
element $W$ such that
\[ \Delta h=W(h\tens 1)W^{-1},\quad Sh=(\eps\tens\id)\circ W^{-1}(h\tens (\
)),\qquad W_{12}W_{13}W_{23}=W_{23}W_{12}.\]
Here $W_{12}=W\tens 1$, $W_{23}=1\tens W$ in $\Lin(H^{\tens
3})\supseteq\Lin(H)^{\tens 3}$ (similarly for $W_{13}$). Let $H^*\subset
\Lin(H)$ by
$\phi\mapsto \phi\la=R^*_\phi$ (the left coregular representation
$R^*_\phi=(\id\tens\phi)\Delta$). Then viewed in this algebra we also have,
where defined,
\[ \Delta \phi=W^{-1}(1\tens\phi)W,\quad S\phi=(\id\tens\phi)\circ W^{-1}((\
)\tens 1)).\]
In the finite-dimensional case the subalgebras $H\subset \Lin(H)$ and
$H^*\subset \Lin(H)$ together generate all of $\Lin(H)$.
\end{theorem}
\proof Note that the set of linear maps $\Lin(H\tens H)$ contains
$\Lin(H)\tens\Lin(H)$ in the standard way but may be larger in the
infinite-dimensional case. Explicitly, $W,W^{-1}$ are defined as such linear
maps $H\tens H\to H\tens H$ by
\eqn{funda}{ W(g\tens h)=\sum g\o\tens g\t h,\qquad W^{-1}(g\tens h)=\sum
g\o\tens (Sg\t)h.}
Here $\Delta g=\sum g\o\tens g\t$ is a standard notation\cite{Swe:hop}. For
brevity, we will omit the $\sum$ signs.  We first verify the identities
\align{(W(h\tens 1)W^{-1})(g\tens g') \nquad&&= W(h\tens 1)\la(g\o\tens
(Sg\t)g) = W(hg\o\tens (Sg\t)g') \\
&& = h\o g\o\o \tens h\t g\o\t (Sg\t) g' = h\o g\tens h\t g' = (\Delta
h)\la(g\tens g'),}
\align{&&\nquad\nquad (W^{-1}(1\tens\phi)W)(g\tens g') =
W^{-1}(1\tens\phi)\la(g\o\tens g\t g') = W^{-1}(g\o\tens g\t\o g'\o)<\phi,g\t\t
g'\t>\\ &&\nquad= g\o\o\tens (Sg\o\t)g\t\o g'\o)<\phi,g\t\t g'\t> = g\o\tens
g'\o<\phi,g\t g'\t> = \phi\o\la g\tens \phi\t\la g'.}
Similarly for the antipodes. As for the equations satisfied by $W$ itself, we
evaluate on $H\tens H\tens H$ as \align{ W_{12}W_{13}W_{23}(g\tens h\tens
f)\nquad&&=W_{12}W_{13}(g\tens h\o\tens h\t f)=W_{12}(g\o\tens h\o\tens g\t h\t
f)\\
&&=g\o\o\tens g\o\t h\o \tens g\t h\t f=g\o\tens g\t\o h\o\tens g\t\t h\t f\\
&&= W_{23}(g\o\tens g\t h\tens f)=W_{23}W_{12}(g\tens h\tens f).}
For the last part we show that every operator $H\to H$ arises by actions $\la$
of $H$ and $H^*$ on $H$, at least when $H$ is finite-dimensional. In this case
every linear operator can be viewed as an element of $H\tens H^*$ acting on $H$
in the usual way by evaluation, namely $(h\tens\phi)(g)= h<\phi,g>$. We have to
represent this by elements of $H,H^*$
acting via $\la$. Indeed, as operators in $\Lin(H)$ we find $(h\tens \phi)=h
(S^{-1}e_a\o)\la (<\phi,e_a\t> f^a\la(\ ))$ where $e_a$ is a basis of $H$ and
$f^a$ a dual basis. Thus $h (S^{-1}e_a\o)\la(<\phi, e_a\t>\tens f^a\la g)=h
(S^{-1}e_a\o)g\o<\phi, e_a\t><f^a,g\t>=h (S^{-1} g\t\o)g\o<\phi, g\t\t>= h
<\phi,g>$ as required. We used coassociativity and the properties of $S^{-1}$
as skew-antipode. \endproof

If $W$ is any invertible operator obeying the { pentagon identity} $
W_{12}W_{13}W_{23}=W_{23}W_{12}$ then it is easy to see that
$\Delta(h)=W(h\tens 1)W^{-1}$ for any operator $h$ will always be coassociative
and an algebra homomorphism, cf\cite{EnoSch:dua}. If we can arrange also for a
counit (typically by restricting our operators to some subalgebra) then this
gives a Hopf algebra. A Hopf algebra of this type is very concrete, being
realised as operators on some vector space, and hence very suitable for
physical applications. On the other hand, it might be thought that only very
special
Hopf algebras could be obtained concretely in this way, but our theorem says
that {\em every} Hopf algebra can be realized concretely by acting on itself.
Even more, it says that both the Hopf algebra and its dual can be realized
concretely at the same time as subalgebras of operators on the same space.

For a quantum-mechanical setting, as well as to make contact with the existing
theory of Hopf-von Neumann and Kac algebras, we consider now the situation when
$H$ is a Hopf $*$-algebra. To do this, it is helpful to cast the last part of
the preceding theorem as a statement about Weyl algebras. If $H,H^*$ are dually
paired Hopf algebras, we define $w(H)$ (the Weyl algebra of $H$) to be the
semidirect product $H\cocross H^*$ where $H^*$ acts on $H$ by the left action
$\la$ above. The multiplication is defined in the linear space $H\tens H^*$ by
\eqn{weyl}{ (h\tens\phi)(g\tens\psi)=\sum hg\o\tens \phi\o\psi <\phi\t,g\t>.}
Both $H,H^*$ are subalgebras. Incidentally, this is symmetric in a certain
sense between $H,H^*$; one has $w(H^*)\isom w(H)^{\rm op}$ at least if $H$ has
a bijective antipode (which we assume). Now, if $H$ is a Hopf $*$-algebra then
so is $H^*$ and $w(H)$ becomes a $*$-algebra by
\eqn{weylstar}{(h\tens\phi)^*=(1\tens\phi^*)(h^*\tens 1).}
This is the only possibility when we bear in mind that $(h\tens\phi)=(h\tens
1)(1\tens\phi)$. It is easy to see that it  indeed gives a $*$-algebra and that
$H,H^*$ are $*$-subalgebras.

\begin{corol} Let $H$ be a finite-dimensional Hopf algebra. Then $\Lin(H)\isom
w(H)$ as algebras. The left hand side has the usual composition of linear maps.
Hence if $H$ is a Hopf $*$-algebra, $\Lin(H)$ becomes a $*$-algebra.
\end{corol}
\proof The first part follows at once from the last part of the theorem,  where
$h\tens \phi$ in $w(H)$ built on $H\tens H^*$ corresponds to $h\la(\phi\la(\
))$ in $\Lin(H)$. Conversely, from the proof there, the element  corresponding
to $h\tens\phi\in \Lin(H)$ (acting in the usual way) is $h (S^{-1}e_a\o)\tens
<\phi,e_a\t> f^a\in w(H)$ built on $H\tens H^*$. Hence $\Lin(H)$ is a
$*$-algebra inherited from the $*$-algebra structure of $w(H)$. Since $H,H^*$
are subalgebras, this is such that $(h\la)^*=h^*\la$ and
$(\phi\la)^*=\phi^*\la$ as operators in $\Lin(H)$. \endproof

\begin{propos} In the setting of Theorem~3.1, suppose that $H$ is a
finite-dimensional Hopf $*$-algebra and let $\Lin(H)$ have the $*$-structure
generated by $H,H^*$ (i.e. from $w(H)$). With respect to this $*$-structure,
the fundamental operator $W$ is unitary,
\[ W^{*\tens *}=W^{-1}.\]
The right-invariant integral $\int$ on $H$  defines a sesquilinear form
$(g,h)=\int g^*h=\overline{(h,g)}$ which is compatible with the $*$-structure
on $\Lin(H)$ induced by $H,H^*$ in the sense
\[ \int (h\la g)^*g'=\int g^*(h^*\la g'),\qquad \int (\phi\la g)^*g'=\int g^*
(\phi^*\la g'),\qquad g,g',h\in H,\ \phi\in H^*.\]
\end{propos}
\proof For the first part we begin by writing $W$ in terms of such elements
from $H,H^*$. Indeed, $W(g\tens g')=g\o\tens g\t g'=g\o<f^a,g\t>\tens
e_ag'=f^a\la g\tens e_a\la g'$, where $e_a$ is a basis of $H$ and $f^a$ is a
dual basis. So $W=f^a\la\tens e_a\la$ and hence , by definition, $W^{*\tens
*}=f^a{}^*\la \tens e_a{}^*\la=f'^a\la\tens  (Se'_a)\la$ where
$e'_a=S^{-1}e_a{}^*,f'^a=f^a{}^*$ are a new mutually dual basis. Dropping the
primes, the action of
the result is $f^a\la g\tens (Se_a)\la g'=g\o<f^a,g\t>\tens (Se_a)g'=g\o\tens
(Sg\t)g'=W^{-1}(g\tens g')$.

For the second part we show that this $*$-structure on $\Lin(H)$ really is an
adjoint operation with respect to $(\ ,\ )$. The right integral is
characterized by $(\id\tens\int)\Delta=1\int$ and is unique up to scale. This
means at once that $(S\int)^*=\int$ since $(S\int)^*$ is also a right integral
and has the same normalization, and this implies that $(\ ,\ )$ is hermitian as
stated.
For the operators $h\la$ we have $(h\la g,g')=\int (h\la
g)^*g'=\int(hg)^*g'=\int g^* h^*g'=\int g^*(h^*\la g')=(g,(h\la)^*g')$ is
automatic. Rather harder is for the operators $\phi\la$,
\align{ &&\nqquad (\phi\la g,h)=\int (\phi\la g)^*h=\int
(g\o)^*\overline{<\phi,g\t>}h=\int g^*\o\overline{<\phi,(g^*\t)^*>}h\\
&&=\int g^*\o<S^{-1}(\phi^*),g^*\t>h=\int g^*\o
h\o<S^{-1}\phi^*\th,g^*\t><(S^{-1}\phi^*\t)\phi^*\o,h\t>\\
&&=\int g^*\o
h\o<(S^{-1}\phi^*\t)\o,g^*\t><(S^{-1}\phi^*\t)\t,h\t><\phi^*\o,h\th>\\
&&=\int g^*\o h\o<S^{-1}\phi^*\t,g^*\t h\t><\phi^*\o,h\th> \\
&&=\int (g^*h\o)\o<S^{-1}\phi^*\t,(g^*h\o)\t><\phi^*\o,h\t>\\
&&= \int g^* h\o <\phi^*,h\t>=\int g^*(\phi^*\la h)=(g,(\phi\la)^*h)
.}
Here the second equality is from the definition of $\phi\la g=g\o<\phi,g\t>$.
The third equality is from the definition of a Hopf $*$-algebra with regard to
$\Delta\circ*$. The fourth equality is from the relation between the $*$
structures in $H,H^*$ and $(S\circ *)^2=\id$. The fifth equality inserts some
factors that collapse to $\eps(h\t)$ via the fact that $S^{-1}$ is a
skew-antipode for $H$. The sixth equality writes the multiplication in $H^*$ in
terms of the
comultiplication in $H$, and uses coassociativity and that $S^{-1}$ is an
anticoalgebra map. The seventh equality writes the comultiplication in $H^*$ in
terms of $H$ to combine some factors, while the eighth equality uses that
$\Delta$ is an algebra homomorphism. By these manipulations we are ready, in
the ninth equality, to use that $\int$ is a right integral, leading to the
required result. \endproof

In a functional-analytic context we can take for $H$ a von-Neumann algebra, and
in place of $\Lin(H)$ we take $B(\CH_\phi)$, the bounded operators on the
Hilbert space ${\CH}_\phi$ determined by a state or weight $\phi$. $H$ is
embedded in this by the GNS construction. With $\phi=\int$  (the integral on
$H$), we have that the $*$-algebra structure on $\Lin(H)$ becomes the usual
adjoint operation $\dagger$ on $B(\CH_\phi)$, so that $W^\dagger=W^{-1}$. The
structure in Theorem~3.1 and Proposition~3.3 is then characteristic of a Kac
algebra\cite{EnoSch:dua}. This was historically limited to $S^2=\id$, but this
is not needed when formulated along the lines above, cf \cite{BaaSka:uni}.
Although we will work algebraically, we keep in mind this Hopf-von Neumann or
Kac algebra setting  when we consider quantum mechanical examples. One
complication in the Hopf-von Neumann or Kac algebra setting  is that the counit
is typically unbounded.

As explained in the introduction, the significance of these algebraic results
for our present purposes is that they give a direct `quantum mechanical'
interpretation of the random walk associated to a Hopf algebra $H$ and a linear
functional $\phi$. In terms of the Markov transition operator we see that
\eqn{MTW}{T_\phi(h)=(\phi\tens \id)\left(W(h\tens 1)W^{-1}\right).}
Here $\phi$ should be thought of as giving the expectation value over the first
copy of $H$ in $H\tens H\subset \Lin(H)\tens\Lin(H)$.
Thus our random walk, considered actively as a Markov process consist of the
following. First, we embed $h\in H$ as $h\tens 1\in H\tens H$. Here the first
copy of $H$ is the algebra of observables at time $t$ and the second copy is
the algebra of observables at time $t+\delta$ (i.e. one step further in time).
Then we evolve $h\tens 1$ by the quantum evolution operator $W$ of the joint
system $H\tens H$. Finally, we take a conditional  expectation value in the
first copy of $H$ to leave us in the second copy at $t+\delta$. This represents
`forgetting' the details of where the system might have been at the now
unobserved time $t$. Thus $T_\phi$ is the quantum step of a quantum random walk
in a reasonably physical way.

\section{Duality, Coentropy and Time-Reversal}

In this section we explore the implications of Hopf algebra duality for the
interpretation of quantum random walks on Hopf algebras. A parallel
observable-state symmetry in the context of quantum-gravity was explored in
\cite{Ma:pla}. The point of \cite{Ma:pla} was that when an algebra of
observables is a Hopf $*$-algebra, then the dual $H^*$ is also a (Hopf)
$*$-algebra. Thus we can, in principle, regard $H^*$ instead as the algebra of
observables of some dual quantum
system. The original observables are now regarded as elements of $H^{**}$ i.e.
linear combinations of states from the dual point of view. Roughly speaking
(and not worrying about positivity) the same expectation value $\phi(h)$ of
observable $h$
in state $\phi$ is regarded from the dual point of view as $h(\phi)$, the
expectation of $\phi$ in state $h$. We gave several examples in
\cite{Ma:pla}\cite{Ma:hop} based on quantum particles on homogeneous spaces.
The classical data was
a pair of groups $(G_1,G_2)$ (the momentum and position groups respectively)
acting on each other in a compatible way. The dual Hopf (von Neumann-Kac)
algebra was of the same type with the roles of $G_1,G_2$ interchanged. In the
dual picture the quantum particle moves along orbits of $G_2$ in $G_1$ instead
of orbits of $G_1$ in $G_2$. Thus, interesting models certainly exist in which
the dual system has just as good a physical interpretation as the original
system and for which the implications of this observable-state symmetry can be
explored. Our considerations will be at a formal algebraic level (there exist
algebraic examples too\cite{Ma:phy}) but the most natural setting that we have
in mind is this Hopf-von Neumann or Kac algebra one.

To explore this duality in the present context we fix $H$ a Hopf $*$-algebra
and $\phi$ a state on it. We have seen that $\phi$ generates a quantum random
walk with transition operator $T_\phi$. Note that there is clearly an arrow of
time built into this interpretation as we successively take each step of the
walk. It can be expressed by the fact that the completely positive operator
$T_\phi$ necessarily increases the entropy of $\phi^n$, as follows.

Firstly, the relative entropy $\CS(\phi,\psi)\le 0$ is defined between two
states $\phi,\psi$ in a standard way. For the algebra of bounded operators on a
Hilbert space, states are of the form $\phi=\sum_i s_i <\phi_i|\
|\phi_i>,\psi=\sum_j r_j <\psi_j|\ |\psi_j>$ (convex linear combinations of
pure states) and
\eqn{relent}{ \CS(\phi,\psi)=\sum_{i,j}(-s_i\log s_i +s_i\log r_j)\vert
<\phi_i|\psi_j>\vert^2.}
This definition extends to abstract von Neumann and $C^*$ algebras as well as
into certain algebraic situations. For recent work see
\cite{Don:rel}\cite{CNT:dyn} and elsewhere. For our purposes below we need only
the standard abstract properties of the entropy without worrying too much about
its detailed definition. Its heuristic interpretation, as discussed in
\cite{BraRob:ope}\cite{Don:rel} is that $e^{\CS(\phi,\psi)}$ is the probability
per unit measurement of the system appearing to be in state $\phi$ when it is
in state $\psi$. In very general terms it measures the ratio of the impurity of
$\phi$ to that of $\psi$. For the entropy of a single state an obvious choice
for us is to take the entropy relative to $\eps$ or the entropy of $\eps$
relative to the state. For our random walk we have
\eqn{entepsrev}{0\ge \CS(\eps,\phi^n)=\CS(\eps^{\tens
n}\circ\Delta^{n-1},\phi^{\tens n}\circ\Delta^{n-1})\ge \CS(\eps^{\tens
n},\phi^{\tens n})}
since $\Delta^{n-1}$ is a completely positive map (being a $*$-algebra
homomorphism) and using (\ref{comult}). There is a similar identity
$\CS(\phi^n,\eps)\ge \CS(\phi^{\tens n},\eps^{\tens n})$ although this is less
interesting because $\CS(\phi,\eps)$ etc tend to be $-\infty$ or $0$ (this is
the case classically). These inequalities give some information about the
relative entropy between $\eps$ and $\phi^n$.  We can also get incremental
information in the form
\eqn{entinc}{ 0\ge \CS(\phi^{n},\phi^{n+1})=\CS(\phi^{n-1}\circ
T_\phi,\phi^{n}\circ T_\phi)\ge \CS(\phi^{n-1},\phi^{n})\ge\cdots \ge
\CS(\eps,\phi)}
since $T_\phi$ is completely positive. This says that as $n$ grows, the states
$\phi^n$ of our random walk change more and more slowly in the sense that the
probability for the system to still appear in state $\phi^{n}$ when it is in
the next $\phi^{n+1}$, increases. It also says that the impurity of $\phi^{n}$,
which is generally less than that of $\phi^{n+1}$, tends towards the latter. In
fact, the reason for this is that as the walk evolves, the states are becoming
more and more dissipated or impure, but this however, is in some sense bounded
by the right integral $\int$ from Proposition~3.3 (which we assume is
positive). To see this we note that
\eqn{TI}{ \int \circ T_\phi=(\phi\tens\int)\Delta=\phi(1)\int=\int}
by the right invariance of $\int$. Hence
\eqn{entint}{0\ge \CS(\phi^{n+1},\int)= \CS(\phi^n\circ
T_\phi,\int)=\CS(\phi^n\circ T_\phi,\int\circ T_\phi)\ge \CS(\phi^n,\int)\ge
\cdots \ge \CS(\eps,\int)}
Here the entropy of $\phi^n$ relative to $\int$ indeed increases with each step
and in this sense the state becomes more and more similar (before any
rescaling) to $\int$. It also becomes more and more impure as its degree of
impurity  approaches the degree of impurity of $\int$ as claimed.  This
integral $\int$ represents a kind of `maximal entropy' or `maximally impure'
state, and our random walk evolves towards it. In the classical case of
functions on a finite group, the integral is precisely the maximally impure
state consisting of the average of all the pure states (which are given by
evaluation at the various group elements), while $\eps$ is a pure state (given
by evaluation at the group identity element). In this case it is easy to
compute that $\CS(\eps,\int)=-\log \vert \Omega\vert$ where $\vert \Omega\vert$
is the order of the group $\Omega$. There is a similar formula for
$\CS(\eps,\int)$ for general Hopf algebras and we see that it provides a lower
bound for $\CS(\phi^n,\int)$ .

The origin of this increase of entropy relative to $\int$ is that whereas the
evolution of the joint system by $W$ in (\ref{MTW}) is invertible, the
conditional expectation given by evaluation  in the state $\phi$ represents
forgetting information. This is the reason that repeated application of
$T_\phi$ is entropy increasing in this way. Moreover, we demonstrated  in
Section~2.2 precisely how this discrete arrow of time leads in a limit to a
more familiar continuous arrow of time.

Now for a fixed $H$ we can also consider $a\in H$ as an element of $H^{**}$,
and hence this induces a random walk on $H^*$ (a Hopf algebra dually paired
with $H$). We need to assume that this $a$ is positive  as a linear functional
on $H^*$,

\begin{defin} An element $a$ in a Hopf $*$-algebra $H$ is {\em copositive} if
$<a,\phi^*\phi>\ge 0$ for all $\phi\in H^*$. It is normalised if $\eps(a)=1$.
\end{defin}
For example, if $a$ is grouplike and anti-self-adjoint in the sense $Sa=a^*$,
then it is copositive. Convex linear combinations
\[ a=\sum_i s_i a_i,\qquad Sa_i=a_i^*,\quad \Delta a_i=a_i\tens a_i\]
of such elements $a_i$ are also copositive. We say that a copositive observable
is pure if it is not the convex linear combination of copositives. Now, if $a$
is a normalized copositive element then the random walk induced on $H^*$ has
transition operator $T_a:H^*\to H^*$. This random walk when viewed actively as
an irreversible evolution among the observables of the dual system (i.e. among
elements of $H^*$) can be referred to $H$ where it appears passively as
evolution of the states of the dual system. This is given by the adjoint
operators $T^*_a:H\to H$. We compute
\eqn{LC}{<\phi, T_a^*(h)>=<T_a(\phi),h>=\sum
<\phi\o,a><\phi\t,h>=<\phi,ah>,\quad {\rm i.e.}\quad T_a^*=L_a:H\to H.}
Thus, the transition operator of the dual random walk corresponds to left
multiplication by $a$ when viewed in terms of the original system. Note that
information is still being lost with each step even in this picture. An
immediate corollary of the remarks in Section~2.3 (at least in the
finite-dimensional case) is

\begin{corol} An element $a\in H$ is copositive iff the operator of
left-multiplication $L_a:H\to H$ is the adjoint of a completely positive map.
\end{corol}

This interpretation of a copositive element $a\in H$ as leading to a  quantum
random walk in the dual, leads us to define the notion of coentropy of an
element of $H$.

\begin{defin} Let $a,b\in H$ be copositive elements. The relative coentropy
$\CS{}^*(a,b)$ is the relative entropy of $a,b$ as linear functionals on $H^*$.
\end{defin}
The familiar properties of entropy now appear in dual form for the coentropy.
For example, $a\circ T_b=ba$ as linear functionals on $H^*$ (see (\ref{LC}))
means that
\eqn{coent}{\CS{}^*(ba,bc)\ge \CS{}^*(a,c)}
for three copositive elements. Likewise, the heuristic interpretation of
$\CS{}^*(a,b)$ is that $e^{\CS{}^*(a,b)}$ should be thought of as the
probability per unit trial of mistaking the observable or random variable $b$
for $a$ when examined on random states of the system. In very general terms it
is the ratio of the impurity of $a$ to that of $b$. An example of a copositive
element is provided by the left integral element {\em in} our Hopf $*$-algebra.
This is an element $\Lambda\in H$ such that $h\Lambda =\eps(h)\Lambda$ for all
$h$, and at least in the finite-dimensional case this exists and is unique up
to scale. It is anti-self-adjoint according to $S\Lambda=\Lambda^*$, and in
nice cases is copositive in the above sense. It corresponds to the right
integral $\int$ on $H^*$ and is maximally impure in the way that we have seen
above for $\int$.

We now consider $H$ equipped with a copositive element $a$. It induces a random
walk in $H^*$, but when referred to $H$ this process is the left multiplication
operator. Thus, from the point of view where $H$ is regarded as a quantum
algebra of observables, the second process appears not as a random walk on $H$
but as what we will call a {\em creation process}. Recall that in any GNS
representation of $H$, the element $1$ in $H$ leads to a vacuum vector $\vert
0>$ and $L_a$ turns this into another vector $a\vert 0>$, created by $a$.
Repeating this for a sequence $a_1,\cdots a_n$ gives the vector
\[ a_na_{n-1}\cdots a_1|0>.\]
To be concrete here, one can keep in mind the GNS representation given by the
integral $\int$ as in Proposition~3.3.
In summary, a geometrical process involving the group structure
(comultiplication) of $H^*$ corresponds from the point of view of $H$ to a
quantum process involving  multiplication in the quantum algebra of
observables. This was the point of view of \cite{Ma:pla} also.

This quantum `creation process' also has a simple classical meaning, at least
if we do not worry about the copositivity
or normalization requirements. In the classical case the algebra of observables
is the algebra of functions on a space $\Omega$. Among them, the characteristic
functions of subsets of $\Omega$ correspond to the observables or random
variables that specify membership of the subsets. A sequence $a_1,a_2,\cdots$
of such characteristic functions is then a creation process that successively
specifies more and more conjunctions as we successively multiply more and more
characteristic functions.
The first step of the creation process is 1 (the identity function), the next
step $a_1.1$ specifies membership of one subset, $a_2.a_1.1$ further specifies
an additional membership, etc. This models then the way that classical concepts
or observables are created as a series of specifications. We have seen that
this elementary process corresponds to a random walk in the dual.

Finally, our comments above about the entropy of a random walk now become
comments about coentropy in a creation process. Using (\ref{coent}), the
analogue of (\ref{entint}) is
\eqn{coentint}{0\ge \CS{}^*(a^{n+1},\Lambda)\ge
\CS{}^*(a^n,\Lambda)\ge\cdots\ge \CS{}^*(1,\Lambda).}
This suggests that for a typical copositive element $a$ the sequence
$1,a,a^2,\cdots$ typically tends to $\Lambda$ (and perhaps converges to
something sensible after rescaling according to a dual central limit theorem).
In general terms it  becomes more and more impure as it approaches $\Lambda$ in
this coentropic sense. We can easily see this phenomenon in the classical case
of the algebra of functions on a (say discrete) group $\Omega$, as follows.
In this case the copositives are convex linear combinations of normalised
traces (characters) of representations of $\Omega$, with the pure copositives
being the traces of the irreducible representations. The normalization is
$\eps(a)=a(e)=1$ where $e$ is the group identity. The integral $\Lambda$ is
given by the Kronecker $\delta$-function at $e$. For Abelian groups we have
$a=\sum_i s_i a_i$ where the $a_i$ are the one-dimensional representations of
the group, i.e. the `plane waves'. If we proceed (with care) to the continuous
case $\Omega=\R$ then the pure copositives are just the pure frequency waves
$a_\omega=e^{\imath \omega(\ )}$. Our assertion that typical copositive
elements give creation processes tending to the integral is then something
familiar. For example, we can take the convex linear combination
\[ a=\h e^{\imath\omega(\ )}+\h e^{-\imath\omega(\ )}=\cos \omega(\ )\]
(this just corresponds to Brownian motion  for the dual random walk as in
(\ref{e6})). It is well-known that squaring such cosine waves introduces higher
harmonics. Moreover, raising to higher and higher powers $a^n$ makes the wave
more and more impure as further harmonics are introduced until we have
something resembling the integral $\Lambda$. This is the maximally impure wave
containing all frequencies, i.e. the $\delta$-function. This is a simple
physical interpretation of (\ref{coentint}) in the classical case.

If we do not worry about the copositivity condition, we can also give another
picture of this general phenomenon in terms of membership of subsets as
mentioned above. For this we consider a non-stationary creation process given
by a family of subsets containing $e$ (so that their characteristic functions
$a_1,a_2,\cdots$ are correctly normalised). These indeed typically tend to
$\{e\}$ (corresponding to $\Lambda$) as the only point definitely in their
joint intersection. This is a second `boolean' interpretation of
(\ref{coentint}).

We are now ready to consider a system $(H,\phi,a)$ consisting of a Hopf
$*$-algebra, a state $\phi$ and a normalized copositive element $a$ in $H$. The
first element induces a random walk in $H$. The second induces a creation
process in $H$.  In terms of these elementary considerations, we can express
our observable-state duality in this context as a `CTP'-type proposition.

\begin{propos} Let $(H,\phi,a)$ be a Hopf $*$-algebra equipped with state
$\phi$ (giving a random walk) and a normalised copositive element $a$ (giving a
creation process). Then  $(H^*,a,\phi)$ has the same interpretation but with
the algebra of observables given by $H^*$ rather than $H$ and the roles of
$a,\phi$ interchanged.

More generally, let $(H,\phi_1,\phi_2,\cdots, \phi_n,a_1,a_2,\cdots,a_m)$ be a
non-stationary random walk with successive distributions $\phi_1,\phi_2,\cdots$
for each step, and a non-stationary creation process successively creating the
vectors $a_1|0>,a_2a_1|0>,\cdots$ . Then $(H^*,a_m,a_{m-1},\cdots,
a_1,\phi_n,\phi_{n-1},\cdots,\phi_1)$ has the same interpretation with the
roles of $H,H^*$, $a,\phi$ interchanged and the order reversed, i.e.
successively applying steps $a_m,a_{m-1},\cdots$ and creating vectors
$\phi_n|\eps>,\phi_{n-1}\phi_n|\eps>,\cdots$. Here the vacuum vector $|\eps>$
in $H^*$ corresponds to the counit $\eps$.
\end{propos}
\proof The first part summarises the discussion above. We saw in (\ref{LC})
that the random walk on $H^*$, when referred to $H$ is the left regular
representation. Similarly with the roles of $H,H^*$ reversed. We are of course
expressing nothing other than the self-duality of the axioms of a Hopf
$*$-algebra. For the second part we have to note carefully the order in
(\ref{Tmulti}). The operator $T$ as a map $T:H^*\to \Lin(H)$  is an
anti-representation. Its adjoint  therefore leads to a map $L:H^*\to\Lin(H^*)$
(the creation process on $H^*)$ which is a representation rather than an
antirepresentation (since taking adjoints reverses the order). Note that this
reversal is not a feature of our conventions but a genuine aspect of the
interpretation of $T_\phi$ as explained in Section~2.3 and of the (quantum)
creation process as discussed above.
\endproof

Thus, the combined random-walk-creation process in $H$ can be viewed equally
well as a combined process in $H^*$, but with the time direction built into the
processes on $H^*$ reversed relative to the processes on $H$. Thus the symmetry
between observables and states when the quantum algebra of observables is a
Hopf algebra\cite{Ma:pla} also involves a time reversal. Another way to see
this reversal is to return to Theorem~3.1. There we see also that the
comultiplications of both $H$ and $H^*$ can be understood as conjugation by a
joint evolution operator. For the former (which leads to the
random walk in $H$) this is by embedding $h$ as $h\tens 1$ and conjugation by
$W(\ )W^{-1}$. For the latter (which leads to a random walk in $H^*$) it is by
embedding $\phi$ as $1\tens\phi$ (hence with a left-right or parity reversal)
and conjugation by $W^{-1}(\ )W$ (hence with the inverse  joint evolution
operator). We note that this  time-reversal phenomenon associated with Hopf
algebra duality is reminiscient of CTP-invariance in particle physics. It
suggests a relationship between Hopf algebra duality and particle-antiparticle
duality which will be developed further elsewhere.

\end{document}